\documentclass[aps,showpacs,twocolumn,showkeys,amsmath,amssymb]{revtex4}
\usepackage{graphicx}% Include figure files
\usepackage{dcolumn}% Align table columns on decimal point
\usepackage{bm}% bold math
\usepackage{mathrsfs}
\usepackage{epsfig}

\begin{document}
%UMD-DOE/ER/40762-508 Nov 16 2011 (w/ Boris Gelman,PRC 85 (2012) 024001)
%UMD-DOE/ER/40762-518 Mar 21 2012

\title{Center symmetry and area laws}

\author{Thomas D. Cohen}
\email{cohen@physics.umd.edu}
\affiliation{Department of Physics and the Maryland Center for Fundamental Physics, University of Maryland, College Park, MD 20742-4111, USA}

\date{\today}

\begin{abstract}
SU($N_c$) gauge theories containing matter fields may be invariant under transformations of some subgroup of the $\mathbb{Z}_{N_c}$ center; the maximum such subgroup is $\mathbb{Z}_{p}$, with $p$ depending on $N_c$ and the representations of the various matter fields in the theory.  Confining SU($N_c$) gauge theories in either 3+1 or 2+1 space-time dimensions and with matter fields in any representation have string tensions for representation $R$ given by $\sigma_R =\sigma_f  \, \, \frac{p_R (p-p_R) \, \, g\left (p_R (p-p_R) \right )}{(p-1) \, \, g(p -1 )} $ with $p_R={n_R \, \rm mod}(p)$, where $\sigma_f  $ is the string tension for the fundamental representation, $g$ is a positive finite function  and $n_R$ is the n-ality of $R$. This implies that  a necessary condition for a theory in this class  to have an area law is invariance of the theory under a nontrivial subgroup of the center.  Significantly, these results depend on $p$ regardless of the value of $N_c$.
\end{abstract}

%\pacs{2}% PACS, the Physics and Astronomy
                             % Classification Scheme.
\maketitle

The nature of confinement in QCD and related gauge theories is quite subtle and remains a subject of considerable interest\cite{footnote}.  QCD lacks an order parameter for confinement. However,  various cousins of QCD have well-defined order parameters. In  SU($N_c$) Yang-Mills theory, the Polyakov loop in Euclidean space serves as an order parameter\cite{Polyakov}.  It is connected to center symmetry:  the Euclidean action is invariant under center transformations while the Polyakov loop is not\cite{Greensite}. The presence of quark fields in QCD spoils the invariance of QCD under center transformations. Yang-Mills theory has another set of indicators of confinement that are lacking in QCD, namely area laws for Wilson loops\cite{Wilson} associated with the various representations of the group.  A Wilson loop for a representation $R$, $\sigma_R$, is characterized by its  string-tension
\begin{equation}
\begin{split} 
\sigma_R & \equiv \lim_{ L \rightarrow \infty \, , \, T \rightarrow \infty } \frac{\log \left( \langle W^R \rangle \right )}{ L T} 
\\
W^R & \equiv {\rm  tr} \left( {\rm P} \,  \exp\left  (i \int_C \sum_i  A_i T_i^R \cdot dx \right )  \right)
\end{split}
\end{equation} 
where $W$ is the Wilson loop operator, $C$ is a closed rectangular path of length $L$ in a spatial direction and $T$ in the temporal direction, P indicates path ordering,  $A_i$ is the $i^{\rm th}$ gluon field and  $T_i^R$ is the $i^{\rm th}$ generator in representation $R$.  A nonzero string tension is the defining characteristic of an area law for the Wilson loop. %Physically, $\sigma_R$ is the slope of the linearly rising potential between infinitely massive color charges in representation, $R$. 
Over the years, there has been considerable interest in the dependence of the string tension on representation for $SU(N_c$ Yang-Mills theory and related theories\cite{ks1,ks2,ks3,ks4,ks5,ks6,ks7,ks8,ks9,ks10,ks11,ks12,ks13}. While, there has been controversy as to the detailed form of the dependance on the representation, with Casimir scaling or a sine law scaling being two popular conjectures,
%\begin{equation}
%\begin{split}
%{\sigma_R} &=  \sigma_f  \, n_R \left (1- \frac{n_R-1}{N_c-1} \right )  \; \; \; {\rm Casimir \,  scaling} \\
%{\sigma_R} &=\sigma_f  \frac{\sin\left(\frac{\pi \, n_R}{N_c}  \right )}{\sin\left(\frac{\pi}{N_c}  \right )}  \; \; \;  \; \; \;  \; \; \;  \; \; \; \; \; \; {\rm sine \, law  \,scaling}  \; ,
%\end{split}
%\end{equation}
%where $\sigma_f$ a the string tension for Wilson loops in the fundamental representation,
 that the ratio of string tensions in different representations  depends only on the n-ality of R and $N_c$, and not on the representation itself appears to be generally accepted. The belief is that the long distance behavior is determined by the formation of a ``k-string'' with $k$ being the n-ality of the representation; gluon screening allows all representations with the same n-ality to connect to the same k-string.    In QCD, quarks spoil the area law, forcing all string tensions to zero: physically, the area law breaks down  in QCD  because quark-antiquark pairs can  screen the color charges\cite{Greensite}.  

Center symmetry and an area law for the Wilson loop have long been associated.~'t Hooft's classic paper\cite{tHooft} introducing the notion of center vortices,  did so  to explain the area law. However, the connection between an area law for the Wilson loop in a gauge theory and center symmetry is subtle since a Wilson loop in an infinite space is neutral under center transformations.  This paper explores the connection between the area law for the Wilson loop and center symmetry, for  SU($N_c$) gauge theories generally {\it i.e.}  beyond pure Yang-Mills.    

To explore the connection, a large class of theories, SU($N_c$) gauge theories  in  3+1 or 2+1 space-time dimensions with  matter fields in all allowable representations, is considered.  In 3+1 dimensions, theories with matter fields in large representations are not ultraviolet complete.   However, in 2+1 dimensions, matter fields in all representations are possible \cite{CS}, greatly enlarging the class.  The connection is explored by studying relations among the string tensions which characterized area laws for various representations and relating these to center symmetry.   This large class is interesting since it contains theories where matter fields spoil some, but not all, of the $\mathbb{Z}_{N_c}$ center symmetry,  retaining invariance under a subgroup $\mathbb{Z}_p$ subgroup of the center.   These  matter fields also affect Wilson loops; properties of which that are sensitive only to $p$,  the amount of center symmetry that survives are of interest.  Recall that a hint of a deep connection between the area law and center symmetry is the fact that in QCD, the same thing which spoils center symmetry also spoils the area law, namely the quarks.  The issue explored here, is whether matter fields that spoil only part of center symmetry,  have effects on area laws that are predictable solely from the amount of center symmetry preserved.  For example: what properties of large Wilson loops are shared by  an SU(12) gauge theory in 2+1 space-time dimensions with quarks in a three-index symmetric representation,  an SU(15) gauge theory in 2+1 dimensions with quarks in both the 12-index antisymmetric representation and 9-index symmetric representation, and a pure Yang-Mills theory  for SU(3) in 3+1 dimension, all of which have a maximum $\mathbb{Z}_3$ symmetry?

The principal result of this paper is that in any confining theory in this class, $\sigma_R$, the string tension in representation $R$ ,  is given by
\begin{equation} \begin{split}
&\sigma_R =\sigma_f  \, \, \frac{x_R \, \, g\left (x_R \right )}{(p-1) \, \, g(p -1 )} \; \; \;{\rm with}   \\&{ x_R=p_R (p-p_R) \; \; \; {\rm where} \; \; \; \; p_R = n_R\; {\rm mod}(p)} \; 
\end{split} \label{Eq:princ}
\end{equation}
 the maximum subgroup of the center under which the theory is invariant, $\mathbb{Z}_p$ determines $p$,  $g(x)$ is a positive finite function on the domain $0 \le  x < p^2/4$ that may depend on the theory, $n_R$ is the n-ality of $R$ and $\sigma_f$ is the string tension for the fundamental representation.  The key point is that the relation depends on $p$, the amount of center symmetry, rather than $N_c$, the number of colors.

This result follows from two properties of SU($N_c$) gauge  theories.  Consider a gauge theory with $m$ fields (gluons plus matter fields), 
with the first field carrying n-ality $n_1$, the second field carrying n-ality $ n_2$ etc.  Property i) is  that the maximum subgroup of the center under which the theory is invariant is $\mathbb{Z}_p$ with 
\begin{equation}
p={\rm gcd}(N_c,n_1, n_2, \cdots , n_m) 
\label{cond}\end{equation} 
where gcd is the greatest common divisor.   Property ii) is that there exists a way to combine fields in the theory into a composite in a representation $R$, if $n_R$, the n-ality of representation, is an integer multiple of $p$ as given Eq.~(\ref{cond}).  Significantly, the same value of $p$  appears in both  properties i) and ii).   Ultimately this relates the amount of center symmetry in a theory to the theory's ability to screen color charge in a given representation. After introducing a few basic ideas, these properties will  be proved and from them the principal result and some corollaries derived.

The n-ality of a representation is the number of boxes in the Young tableau specifying the representation---modulo $N_c$\cite{Greensite}.  Thus, the Clebsch-Gordan decomposition of  the product of operators in two representations with n-aility $n_1$ and $n_2$ contains only representations with n-anlity equal to $ (n_1+n_2)\;  {\rm mod}(N_c)$.  Any representation with fixed n-ality  can be obtained from any other representation with the same n-ality by combining it with some number of adjoint representations using appropriate Clebsch-Gordan coefficients.  Thus by adding gluons to a combination of fields in a given representation, a combination of fields in any representation with the same n-ality  can be obtained.

The center group associated with a Lie group contains those elements of the Lie group which commute with all elements of the group\cite{Georgi}.  For  SU($N_c$),  the center is $\mathbb{Z}_{N_c}$ and contains $N_c$  elements given by are $C_j=z_j {\bf 1}$
where ${\bf 1}$ is the $N_c \times N_c$ identity matrix and $z_j=\exp \left (i \frac{2 \pi j}{ N_c} \right )$ with  $j=0,1,2, \ldots N_c-1 $.
Center invariance for a gauge theory has a relative simple  formulation on the lattice\cite{Greensite} but the connection between that formulation and the continuum is a bit subtle.  Here the analysis is based on an equivalent formulation\cite{HW} directly based on the  continuum version of the theory  in Euclidean space.  The formulation depends on the space having a finite extent,  $\beta$, in the temporal direction and with periodic  boundary conditions  for   bosons  and anti-periodic  for fermions ones.  This setup corresponding to working at finite temperature\cite{Kapusta}.  Zero temperature physics can be studied by taking the zero temperature limit at the end of the problem.  A center transformation on the gauge field has the following form:
\begin{align}
& A_\mu \rightarrow A'_\mu  \equiv \Omega A_\mu \Omega^\dagger - g \Omega \partial_\mu \Omega^\dagger \label{trans}  \\
& {\rm with} \; \;  \Omega(\vec{x},\beta+t)  = C \Omega(\vec{x},t) \label{bc}
\end{align}
where $\Omega$ is an element of of the gauge group at any point in space time,  ${\bf 1}$ is the identity element, $g$ is the gauge coupling and $C$ is a nontrivial element of the center.  Equation (\ref{trans}) is of the form of a local gauge transformation and leaves the Yang-Mills Lagrangian density invariant at every space-time point.   It is not a true gauge transformation since in a gauge transformation $\Omega$  satisfies periodic boundary conditions: $\Omega(\vec{x},\beta+t)   = \Omega(\vec{x},t)$.  However, while $\Omega$ is not periodic, if $A_\mu$ satisfies periodic boundary conditions with $A_\mu(\vec{x},0)=A_\mu(\vec{x},\beta)$ then so does $A'_\mu$.   For pure Yang-Mills theory, the only fields are the gluons whose boundary conditions are preserved by all center transformations. The transformations are thus  allowable within the theory and leave the action invariant;  Yang-Mills theory is center invariant.

 For theories containing matter fields, the matter transforms  under  center transformations in  the same way as  under gauge transformations:  quarks in the fundamental representation $q \rightarrow q' \equiv \Omega q$ while quarks in the adjoint representation transform according to $q^{\rm adj} \rightarrow q'^{\rm adj} \equiv \Omega q^{\rm adj} \Omega^\dagger$, etc. If such a transformation is allowable given the boundary conditions, it leaves the action invariant as it is of the form of a gauge transformation.  However, it need not be allowable.  Consider a theory  containing a field $\psi$ with n-ality   $n_\psi$.   Under center transformations associated with a particular element of the center $\mathbb{Z}_j=z_j {\bf 1}$ in which $\psi \rightarrow  \psi'$,  the following identity must hold:
 \begin{equation}
 \frac{\psi(\vec{x},t)}{\psi(\vec{x},t+\beta)}= z_j^{n_\psi} \frac{\psi'(\vec{x},t)}{\psi'(\vec{x},t+\beta)} \; .
 \label{n-alitybc}
 \end{equation}
From the boundary conditions, ${\psi_n(\vec{x},t)}/{\psi_n(\vec{x},t+\beta)}$ is fixed to be $\pm1$ depending on whether the field is a bosonic or fermionic.   If $\psi$ satisfies the boundary conditions, then $\psi'$ only satisfies the boundary conditions if 
\begin{equation}
z_j^{n_\psi}= e^{i \frac{2 \pi j n_\psi}{ N_c}}=1 \; .  \label{unity}
\end{equation}
Unless Eq.~(\ref{unity}) holds for all fields in the theory, the center  transformation is not allowable.   If $n_\psi=0$ for all matter fields, as happens if they are in the the adjoint representation, then Eq.~(\ref{unity}) is always satisfied;  the  theory is center symmetric.   If the theory contains a field with $n_\psi = 1$, as happens in a theory with quarks in the fundamental representation, then  Eq.~(\ref{unity}) is {\it never} satisfied except for $j=0$, which is trivial since the ``transformation'' is just the identity.    Such theories are not center symmetric.

If the theory has matter fields with n-ality different from zero or one, the situation is more interesting.  
%Consider for example, a variant of QCD in which the matter fields are in the two-index symmetric (S) or antisymmetric (AS) representation, {\it i.e. fields with n-ality 2}.    Such theories have received considerable attention over the past decade\cite{**}.     If  $N_c$ is even Eq.~(\ref{unity})  $j=N_c/2$ is an allowable nontrivial center transformation that preserves the boundary condition.   Clearly, it is the only allowable nontrivial  center transformation: the boundary conditions imply that the only possible values of $z$ are $\pm 1$.  The theory, while lacking  the full $\mathbb{Z}_{N_c}$ center symmetry is invariant under  a  nontrivial subgroup, namely $\mathbb{Z}_2$.    When $N_c$ is odd; in that case the theory is not invariant under any nontrivial center transformations.  Again the boundary conditions  imply that the only possible nontrivial element of the center is $- \bf{1}$ but $- \bf{1}$n is not an element of the center for odd $N_c.   The fact that $N_c$ even and odd are qualitatively different is reminiscent of SO($N_c$) Yang-Mills theory which is invariant under $\mathbb{Z}_2$ center transformation for $N_c$ even and has no such symmetry for $N_c$ odd\cite{***}. 
First consider a theory with one type of matter field with n-ality, $n_\psi$.   If  $n_\psi$ and $N_c$  are relatively prime, then no values of $j$ other than zero satisfy Eq.~(\ref{unity}).  Conversely, if $n_\psi$ and $N_c$ are not relatively prime, then Eq.~(\ref{unity}) is satisfied if, and only if, $j= l N_c/p$, where $p={\rm gcd}(N_c,n_\psi)$ (with gcd indicating greatest common divisor) and $l$ is a nonnegative  integer less than $p$: transformations associated with the $\mathbb{Z}_p$  subgroup of the  $\mathbb{Z}_{N_c}$ center group are allowable.    In the most general case, there are  multiple fields in the theory, gluons and  some number of matter fields.   The first matter field  carries n-ality $n_1$; the second field carries n-ality $ n_2$; etc.   For a would-be center transformation to be allowable; it must separately preserve the boundary conditions for  {\it all}  of the matter fields as well as for the gluons.  Thus the set of allowable transformations is for the group whose elements are simultaneously elements of $\mathbb{Z}_{p_1}$, $\mathbb{Z}_{p_2}$ , $\ldots$ up through $\mathbb{Z}_{p_m}$ with $p_i={\rm gcd}(N_c,n_i)$.  This group is $\mathbb{Z}_p$ with $p$ given by Eq.~(\ref{cond}).

Having established  property i), consider property ii).   Even if a theory has no matter field in  representation $R$, it may still contain  matter in that representation.   Multiple fields in a theory  can combine into a configuration in representation $R$.   For example,  in Yang-Mills theory for the exceptional group $G_2$,  three gluon fields (in the adjoint ) can combine into the fundamental\cite{G2}. An $SU(N_c)$ gauge theory in 3+1 or 2+1 space-time dimensions will have  matter in representation $R$ if, and only if, matter fields can be combined to yield a representation with n-ality $n_R$.  
%(It should be noted in passing, that this does not apply to 1+1 dimensions unless the theory has adjoint matter as there are no dynamical gluon in 1+1 dimension)
%
% Consider as a specific example, the case where  the matter fields are quarks with n-ality two, such as QCD(S) or QCD(AS)\cite{**}.  One can combine together any number of quarks and antiquarks and the n-abilities of each contribute additively.  Thus, the possible n-abilities, $n$ are $n= 2 l \, {\rm mod}(N_c)$ where $l$ is an integer (which can be negative since antiquarks have n-ality of $N_c-2$ which is equivalent to -2).   When $N_c$ is even the n-abilities reached by combining quarks and/or antiquarks are $0, 2, 4, \ldots N_c-2$ and are all even.  They do not include n-ality unity and hence do not include the any representation from which the fundamental representation can be reached by the addition of gluons.  In contrast, when $N_c$ is odd, the possible n-abilities are $0,1,2, \ldots N_c-1$ and does include unity.  Thus for $N_c$ even the fields in the theory fields cannot combine into the fundamental representation while for $N_c$ odd they can.  Note that for $N_c$ even these theories possess a $\mathbb{Z}_2$ center symmetry while for $N_c$ odd there is no nontrivial center symmetry.  As shown below this pattern is general.
%
Consider, the most general  SU($N_c$) theory  which has   $m$ fields carrying n-ality $n_1, n_2, \ldots n_m$ and ask whether fields in the theory can be combined into representation $R$.  The issue amounts to whether fields can be combined into $n_R$ where  the n-alities add when fields are combine together.  Thus, the possible n-alities are 
$n=\sum_{i=0}^m l_i  \,n_i \; {\rm mod}(N_c) $  where $l_i$ are integers.  This implies that a sufficient condition for fields to combine into representation $R$ is that there there exists a set of integer $l_i$ which satisfy the equation
\begin{equation}
\sum_{i=0}^m l_i  \,n_i \; {\rm mod}(N_c) =n_R\; .
\label{sum}\end{equation}
However, an elementary result from number theory, a generalization of B\'ezout's identity\cite{NT}, implies that if 
\begin{equation}
r \, {\rm gcd}(n_1, n_2, \ldots,n_m,N_c) ={n_R} 
\end{equation}
where $r$ is a positive integer, then there exists a set of integers $l_1,l_2, \ldots, l_m,L$ such that 
$L N_c + \sum_{i=0}^m l_i \, m_i =n_R$
which is equivalent to Eq.~(\ref{sum}).  Thus, fields in the theory can be combined into representation $R$ if  $n_R = r p$ for some positive integer $r$ where $p$ is given by Eq.~(\ref{cond}),   establishing property ii). Superficially properties i) and ii)  deal with quite different things: the amount of center symmetry and the representations of matter which the theory possesses.  However, they are connected in that they both depend  on $p$ as given by Eq.~(\ref{cond}); the connection is number theoretic in origin.

The physical picture for area laws in various representations in Yang-Mills theory is the k-string\cite{Greensite,ks1,ks2,ks3,ks4,ks5,ks6,ks7,ks8,ks9,ks10,ks11,ks12,ks13}.  This is the lowest lying flux tube configuration for a color source carrying n-ality $k$.  Only the n-ality, $k$, matters as gluon screening can shift the representation of the color charge to one with the same n-ality, allowing the system to relax to the lowest energy flux tube with n-ality k.  (Note, that the argument that only n-ality matters is valid in 3+1 and 2+1 space-time dimensions but not in 1+1  dimension where there are no dynamical gluons.  However, it applies to theories in 1+1 dimensions which have matter fields in the adjoint.)  Moreover,  the string tension  for representations with n-ality $k$ and those with  $N_c-k$ are identical since $N_c-k$ is equivalent to $-k$ and simply amounts to switching all color charges to their conjugates ({\it eg.}~the fundamental to the anti-fundamental) which clearly couple to the same k-string.  The string tension for n-ality zero representations vanishes  since in these cases the color charge can be fully screened.   

One expects the k-string picture to be valid beyond Yang-Mills and to hold for confining theories in the large class of theories  considered here. Clearly in this larger class, $\sigma_R=0$ if $n_R=0$ since in this case, the color charge can be fully screened, just as in pure Yang-Mills.  The principal physics difference between this larger class and Yang-Mills is that matter fields can also screen color sources.   As a consequence:
\begin{itemize}
 \item $\sigma_R$, the string tension of representation depends only on $n_R \; {\rm mod}(p)$ where $p$ is given in Eq.~(\ref{cond}).  This follows from Property ii) which means that screening can change the n-ality by an integral multiple on $p$.  Representations whose n-ality differs by an integral multiple of $p$ couple to the same $k$ string and have the same same string tension.
 \item $\sigma_R =\sigma_{R'}$ if $ p_R=(p-p_R')$.  This follows from  the fact that $N_c$ is an integer multiple of $p$, the fact that string tension only depends on the $n_R \; {\rm mod}(p)$ and from charge conjugation which implies that the string tension for representations with n-ality $k$ and $N_c-k$.
 \end{itemize}
 These  facts are fully encoded in Eq.~\ref{Eq:princ} provided $g(x)$ is positive and finite and $p$  is given  by Eq.~(\ref{cond}).  Since, property i) implies that $\mathbb{Z}_p$ is the maximum subgroup of the center for the theory,  where $p$ is also given in Eq.~(\ref{Eq:princ}),  the principal result of this paper relating the maximum subgroup of the center to properties of the string tension has been established.  It is important to stress that Eq.~(\ref{Eq:princ}) depends on the ``p-ality'', {\it i.e.}~the n-ality ${\rm mod}(p)$.  In effect,  things  depend on $p$ regardless of $N_c$: the center group under which the theory is invariant, rather than $N_c$,  determines which string tensions are identical. 
 
Four significant corollaries follow from this result:
 \begin{enumerate}
 \item For theories with a maximum subgroup of $\mathbb{Z}_2$ or $\mathbb{Z}_3$, all representations which have the a nonzero string tension have the same string tension.  This follows from the structure of $x_R$; for $p=2$ all representations have  either $x_R=0$ or $x_R=1$, while for $p=3$, $x_R=0$ or $x_R=2$. \label{cor1}
\item A necessary condition for any theory in the class to have a nonzero string tension for representation $R$ is for $n_R$ not to be an integer multiple of $p$.  This follows from the fact that $x=0$ whenever $n_R$ is an integer multiple of $p$. \label{cor2}
\item A necessary condition for a theory in the class to have an area law for Wilson loops for the fundamental representation is for the theory to be be invariant under a nontrivial subgroup of the center.  This follows since the trivial subgroup has $p=1$ which implies that $x_R=0$, and hence a vanishing string tension, for all representations.\label{cor3}
\item A necessary condition for any theory in the class to have an area law for Wilson loops for all representations with non-zero n-ality is for the theory to be be invariant under the full $\mathbb{Z}_{N_c}$ center group.  This follows directly from corollary \ref{cor2}. \label{cor4}
\end{enumerate}
Again, it should be stressed that all of these corollaries depend the size of the center group rather than $N_c$.  Corollary \ref{cor3} is particularly significant.  It indicates that the connection between area laws and center symmetry is profound---invariance under some nontrivial center transformations is necessary for area laws to exist in this large class of gauge theories.   While this is in accord with the prevailing ``folklore'' of the field, it is gratifying to see formally both that it holds quite generally, and why.  

Corollary \ref{cor1} shows that theories with $\mathbb{Z}_2$ or $\mathbb{Z}_3$ as the maximum subgroup of the center have the ratio of the string tension in any representation to that of the fundamental fixed to be unity or zero.  Remarkably this is regardless of any  other details of the theory including $N_c$, the dimension of space-time or the precise matter content of the theory.  It is interesting to speculate on whether theories with $p>3$ also have universal behavior with $\sigma_R/\sigma_f$, fixed entirely by $p_R$ and $p$ independently of all other details.  If true, the dependence of the ratio on $p_R $ and $p$ must be the same as the dependence on $n_R$ and $N_c$ respectively as in super Yang-Mills theory, since that is in the class.  This would imply a sine law as SYM is known to have this behavior\cite{Greensite} and would fix $g$ to be $g(x) = A  \cos \left ( \frac{\pi}{2} \, \sqrt{1-\frac{4 x}{p} } \right )$ where $A$ is an arbitrary constant.    However, at present one does not know whether $g(x_R)$ is universal.  Perhaps, future numerical lattice studies can shed light on the issue.  \\
 \begin{acknowledgments}
The support of U.S. Department of Energy  is gratefully acknowledged.   The author acknowledges Prabal Adikarhi, Paulo Bedaque, Evan Berkowitz, Michael Cohen and Jeff Greensite for useful discussions.  
\end{acknowledgments}

%\bibliographystyle{unsrt}
%\bibliography{Center}

\begin{thebibliography}{99}
\bibitem{footnote}{For a pedagogical review of many of the issues which arise in this paper see ref.~\cite{Greensite}.}
\end{thebibliography}

\end{document}